\documentclass[12pt]{article}

\usepackage{graphicx}
\usepackage{amsfonts}
\usepackage{amsthm}
\usepackage{amsmath}
\usepackage[top=3cm, bottom=3.25cm, left=2.75cm, right=2.75cm]{geometry}

\newcommand{\Section}[1]{\section{#1} \setcounter{equation}{0}}
\newcommand{\beq}{\begin{equation}}
\newcommand{\eeq}[1]{\label{#1}\end{equation}}
\newcommand{\ber}{\begin{eqnarray}}
\newcommand{\eer}[1]{\label{#1}\end{eqnarray}}

\numberwithin{equation}{section}

\newcommand{\bbD}[1]{\mathbb{D}_{#1}}
\newcommand{\bbDB}[1]{\bar{\mathbb{D}}_{#1}}

\newcommand{\bbX}[1]{\mathbb{X}_{#1}}
\newcommand{\bbXB}[1]{\bar{\mathbb{X}}_{#1}}

\def\+{{+\!\!\!+}}

\newcommand{\kah}{K\"ahler~}
\newcommand{\pa}[1]{\partial_{#1}}

\begin{document}
\renewcommand{\theequation}{\thesection.\arabic{equation}}
\setcounter{page}{0}
\thispagestyle{empty}
\begin{flushright} \small
UUITP-07/09 \\
\end{flushright}

\smallskip
\begin{center}
 \LARGE
{\bf   Pseudo-hyperk\"ahler Geometry and  Generalized K\"ahler Geometry}
\\[12mm]
 \normalsize
{\bf M.~G\"oteman and U.~Lindstr\"om}, \\[8mm]
 {\small\it
Department of Physics and Astronomy,\\
Division for Theoretical Physics,\\
Uppsala University, \\ Box 803, SE-751 08 Uppsala, Sweden}
\end{center}
\vspace{10mm} \centerline{\bfseries Abstract} \bigskip

\noindent We discuss the conditions for additional supersymmetry and twisted supersymmetry in $N=(2,2)$  supersymmetric non-linear sigma models described by one left and one right semi-chiral superfield and carrying a pair of non-commuting complex structures. Focus is on linear non-manifest  transformations of these fields that have an algebra that closes off-shell. 
We find that additional linear supersymmetry has no interesting solution, whereas additional linear \emph{twisted} supersymmetry has solutions with interesting geometrical properties. 
We solve the conditions for invariance of the action and show that 
 these solutions correspond to a bi-hermitian metric of signature $(2,2)$ and a pseudo-hyperk\"ahler geometry of the target space. 
\footnote{
{\bf Keywords:} supersymmetry, semi-chiral fields, sigma models, pseudo-hyperk\"ahler geometry.\\
{\bf Mathematics Subject Classification (2010):} 32C11, 53B30, 53B50, 53C26, 58A50, 81T60, 83E30.}

\eject
\Section{Introduction}
The geometry of the target space of supersymmetric non-linear sigma models is dictated by the number of supersymmetries.  Investigating the conditions under which it is possible to add extra, non-manifest supersymmetries to a sigma model has been a very direct route to finding new and interesting results in complex geometry. In two dimensions
it has led to a complete description\footnote{Away from irregular points.} of generalized K\"ahler geometry (GKG) \cite{Gualtieri:2003dx}. It may be described in terms of a generalized potential $K(\phi,\bar\phi, \chi, \bar \chi, \bbX{L,R}, \bbXB{L,R})$ which depends on chiral $\phi$, twisted chiral $\chi$ and left and right semi-chiral $\bbX{L,R}$, $N=(2,2)$ superfields \cite{Lindstrom:2005zr}. 

The special case of generalized hyperk\"ahler geometry is perhaps less well studied, but a description of additional supersymmetries in purely semi-chiral models was treated already in \cite{Lindstrom:1994mw}. The models described there contain additional $N=(2,2)$ superfields  that are $N=(4,4)$ auxiliaries. Below we describe models with $N=(4,4)$ (twisted) supersymmetry that closes off-shell without such auxiliary fields.

The target space metric for GKG is positive definite, but the development in our understanding of  GKG  has a natural extension to the case of an indefinite (generalized) metric  \cite{DavidovGrantcharovMushkarov'08}. 
In particular, metrics of neutral signature have received increasing attention \cite{Law'92}, \cite{Matsushita'91}, \cite{AFISUV}, \cite{Dunajski:2006mk}, partly because it has been shown that they arise naturally in the context of string theory \cite{Ooguri:1990ww}, \cite{Hull:1997kk}, \cite{AbouZeid:1999em}.
The neutral metrics bear some resemblance to Riemannian metrics, which distinguishes it from other metrics of indefinite signatures. 

In this paper we restrict to four dimensional target space and find that additional supersymmetry cannot be imposed. However, we find a class of interesting solutions with additional \emph{twisted} supersymmetry. 
After describing the $N=(4,4)$ twisted supersymmetry we present the pertinent mathematical background for the neutral hypercomplex structures and then show how a class of such structures arises from potentials in our sigma model setting.

\section{Ansatz}

Consider the generalized \kah potential $K(\bbX{L,R}, \bbXB{L,R})$ for the semi-chiral $N=(2,2)$ superfields $\bbX{L,R}$ satisfying
\ber
\bbDB{+}\bbX{L}=0, \quad \bbDB{-}\bbX{R}=0,
\eer{semis}
where the supersymmetry algebra is
\ber
\{\bbD{+},\bbDB{+}\}=i\pa{\+},\quad \{\bbD{-},\bbDB{-}\}=i\pa{=}.
\eer{salg}
The action 
\ber
S=\int K(\bbX{L,R}, \bbXB{L,R})
\eer{action}
has manifest $N=(2,2)$ supersymmetry.\footnote{Such actions may describe target space geometries with definite or indefinite signature. We conjecture that many of the properties of GKG, such as the existence of a generalized potential, hold for arbitrary signature bi-hermitian geometries.}
One may ask under which conditions the action (\ref{action}) has additional non-manifest symmetries to make it $N=(4,4)$ supersymmetric or twisted supersymmetric.

Supersymmetry can be generalized to twisted supersymmetry \cite{Hull:1997kk}, \cite{AbouZeid:1999em}, where some of the generators close to a pseudo-supersymmetry,
\beq
	\{Q^I, Q^J\} = 2\eta^{IJ}P, \quad \eta^{IJ}=\left(
	\begin{array}{cc}
		1_{p} & 0 \\ 0 & -1_{q}
	\end{array} \right),
\eeq{twistedsusy} 
where $P$ is the translation operators, $p+q=r$ and $I=1,\dots r$. 

In this note we limit the study to four-dimensional target space, where we have only one set of left and right semi-chiral fields and also restrict the additional transformations to be linear in those fields. The general question under which conditions the semi-chiral fields admit a $N=(4,4)$ (twisted) supersymmetry in arbitrary dimension is addressed in a separate paper \cite{workinprogress}.

A general linear transformation that preserves the chirality of the fields reads
\ber
	\delta \mathbb{X}_L
	&=& \phantom{-}i\bar\epsilon^+\bar{\mathbb{D}}_+(\varepsilon \bar{\mathbb{X}}_L + b\mathbb{X}_R+c\bar{\mathbb{X}}_R) + i\kappa \bar\epsilon^- \bar{\mathbb{D}}_- \mathbb{X}_L - i \lambda \epsilon^-\mathbb{D}_- \mathbb{X}_L \nonumber \\
	\delta \bar{\mathbb{X}}_L
	&=& -i\epsilon^+ {\mathbb{D}}_+(\bar\varepsilon {\mathbb{X}}_L + \bar b\bar{\mathbb{X}}_R+\bar c {\mathbb{X}}_R) - i\bar\kappa \epsilon^- {\mathbb{D}}_- \bar{\mathbb{X}}_L + i \bar\lambda \bar\epsilon^- \bar{\mathbb{D}}_- \bar{\mathbb{X}}_L \nonumber \\
\delta \mathbb{X}_R
	&=& \phantom{-}i\bar\epsilon^-\bar{\mathbb{D}}_-(\tilde\varepsilon \bar{\mathbb{X}}_R + \tilde b\mathbb{X}_L+\tilde c\bar{\mathbb{X}}_L) + i\tilde \kappa \bar\epsilon^+\bar{\mathbb{D}}_+ \mathbb{X}_R - i\tilde\lambda \epsilon^+\mathbb{D}_+ \mathbb{X}_R\nonumber\\
	\delta \bar{\mathbb{X}}_R
	&=& -i\epsilon^- {\mathbb{D}}_-(\bar{\tilde\varepsilon} {\mathbb{X}}_R + \bar{\tilde b}\bar{\mathbb{X}}_L+\bar{\tilde c}{\mathbb{X}}_L) - i\bar{\tilde \kappa} \epsilon^+ {\mathbb{D}}_+ \bar{\mathbb{X}}_R + i\bar{\tilde\lambda} \bar{\epsilon}^+\bar{\mathbb{D}}_+ \bar{\mathbb{X}}_R.
\eer{fullansatz}
We now ask whether this ansatz can close to a supersymmetry or a twisted supersymmetry, and whether the transformations keep the  action (\ref{action}) invariant.
\subsection{Supersymmetry}
The transformations (\ref{fullansatz}) close to a supersymmetry algebra
\begin{equation}
	[\delta(\epsilon_1), \delta(\epsilon_2)]\mathbb{X} = i\bar{\epsilon}^\pm_{[2} \epsilon^{\pm}_{1]} \partial_{\pm\!\!\pm} \mathbb{X}
\end{equation}
only if the left- and right sectors decouple. This means that the left semi-chiral field possesses only a left-going supersymmetry, and the right semi-chiral field a right-going. The action (\ref{action}) is invariant under the supersymmetry
\begin{eqnarray}
	\delta \mathbb{X}_L
	&=& i\kappa \bar\epsilon^- \bar{\mathbb{D}}_- \mathbb{X}_L  + \tfrac{i}{\kappa} \epsilon^-\mathbb{D}_- \mathbb{X}_L \nonumber \\
	\delta \mathbb{X}_R
	&=& i\tilde \kappa \bar\epsilon^+\bar{\mathbb{D}}_+ \mathbb{X}_R + \tfrac{i}{\tilde{\kappa}} \epsilon^+\mathbb{D}_+ \mathbb{X}_R
\end{eqnarray}
if and only if the potential $K(\mathbb{X}_L, \bar{\mathbb{X}}_L, \mathbb{X}_R, \bar{\mathbb{X}}_R)$ is linear in all the fields. But since the sigma model (\ref{action}) will vanish for any linear potential, no solution for additional supersymmetry exists. 

\subsection{Twisted supersymmetry}
On the other hand, the transformations can close to a \emph{pseudo}-supersymmetry
\beq
	[\delta(\epsilon_1), \delta(\epsilon_2)]\mathbb{X} = - i\bar{\epsilon}^\pm_{[2} \epsilon^{\pm}_{1]} \partial_{\pm\!\!\pm} \mathbb{X}
\eeq{pseudosusy}
for a larger class of solutions, possessing interesting geometric properties. Closing the pseudo-supersymmetry for the transformations (\ref{fullansatz}), most of the parameters can be solved for. Further, rescaling the fields and transformation parameters in a manner compatible with R-symmetry, the complex $\kappa$ will be the only free parameter. 
The action is invariant under the twisted supersymmetry transformations\,\footnote{The lack of symmetry between the transformations of the left and right fields may seem puzzling but is just an artifact of our choice of rescalings. Symmetric choices are possible.}
\ber
\delta \bbX{L}&=&
i\bar\epsilon^+\bbDB{+}(
\bbXB{L} +\bbX{R}+\tfrac{1}
{\kappa}\bbXB{R})
+i\kappa\bar\epsilon^-\bbDB{-}\bbX{L}-\tfrac{i}{\kappa}\epsilon^-\bbD{-}\bbX{L},\cr
\delta \bbX{R}&=&
i\bar\epsilon^-\bbDB{-}(
\bbXB{R} -
{(\kappa \bar\kappa-1)}
\bbX{L}+
\tfrac{\kappa \bar\kappa-1}{\bar\kappa}
\bbXB{L})
-i
\bar\kappa \bar\epsilon^+\bbDB{+}\bbX{R} +
\tfrac{i}{\bar\kappa} \epsilon^+\bbD{+}\bbX{R},
\eer{lintfs}
provided that $K$ satisfies two complex partial differential equations
\ber
	K_{1\bar 1} - K_{12} - \bar\kappa K_{\bar 1 2} &=& 0,\cr
	(\kappa \bar\kappa -1)K_{2\bar 2} + K_{12} - \kappa K_{1\bar 2} &=& 0.
\eer{invcndn}
The indices $1$ and $2$ denote the partial derivative w.r.t. the left semi-chiral and the right semi-chiral field, respectively.
The system (\ref{invcndn}) may be solved by separating  variables to give a two-parameter family of solutions
\ber
K=F(y) + \bar F(\bar y), \quad y= \alpha \bbX{L}+\beta \bbXB{L} +\gamma \bbX{R} +\delta \bbXB{R},
\eer{50}
where
\ber
\gamma = 
\frac{\alpha\beta}{\alpha + 
\bar\kappa \beta}, \quad
\delta =  
\frac{\alpha\beta}{
\kappa\alpha + \beta}.
\eer{51}
The reason that the solution still depends on two parameters is that the two complex equations in (\ref{invcndn}) have the same imaginary part.
For the model to describe bi-hermitian geometry,\footnote{This condition stems from the fact that $K$ is a generating function for certain symplectomorphisms \cite{Lindstrom:2005zr}. Alternatively, the conditions is needed to integrate out the $(1,1)$ auxiliary fields.}
\beq
\det K_{LR}\ne 0,
\eeq{detcnd1}
where
\ber
K_{LR}\equiv \left(
	\begin{array}{cc}
	K_{12} & K_{1\bar 2} \\
	K_{\bar 1 2} & K_{\bar 1 \bar 2} \\
	\end{array}
	\right).
\eer{Kmatrix}
For the solution (\ref{50}), this condition is equivalent to
\beq
(|\alpha|^2-|\beta|^2)(|\gamma|^2-|\delta|^2)\ne 0.
\eeq{detcnd}
From the linearity of the conditions (\ref{invcndn}), the solution integrated over the free parameters is again a solution,
\ber
\int d\alpha d\beta K(\alpha,\beta; \alpha \bbX{L}+\beta \bbXB{L} +\gamma \bbX{R} +\delta \bbXB{R}).
\eer{52}
\section{Neutral hypercomplex structures}
\label{neutral_hypercomplex_structures}
Consider a smooth $4n$-dimensional manifold $\mathcal M$ with three real endomorphisms $I,S,T : T\mathcal M \hookleftarrow$. Then $(\mathcal M, I, S,T)$ is called a \emph{pseudo-hypercomplex} or \emph{neutral hypercomplex} manifold if the following conditions are fulfilled \cite{Dunajski:2001vt},\cite{Libermann'54}.
\begin{enumerate}
	\item[i.] $(I,S,T)$ satisfy the algebra of split quaternions,
	\begin{equation}	
	-I^2=S^2=T^2=1,\quad IS=T=-SI.
	\end{equation}
	\item[ii.] $(I,S,T)$ are all integrable.\footnote{A sufficient condition is that two of the three structures are integrable. The integrability of the third structure is then implicit \cite{IvanovZamkovoy'03}.} This is equivalent to the vanishing of the Nijenhuis tensor, i.e., if $A\in (I,S,T)$  then
	\begin{equation}
	N(X,Y)=A^2[X,Y] - A[ AX, Y] - A[X,AY] + [AX,AY] = 0
	\end{equation}
	for arbitrary vectors $X,Y \in T\mathcal M$.
\end{enumerate}
Any neutral hypercomplex structure admits a unique torsion-free connection, referred to as the \emph{Obata connection}, such that \cite{IvanovZamkovoy'03}, \cite{AndradaSalamon'03}
\beq
	\nabla I=\nabla S=\nabla T=0.
\eeq{obata}

For a neutral hypercomplex structure $(\mathcal M,I,S,T)$ with a metric $g$, we call $(\mathcal M, I, S, T, g)$ a \emph{neutral hyperhermitian} structure if and only if
\beq
	g(IX,IY)=-g(SX,SY)=-g(TX,TY)=g(X,Y)
\eeq{hermiticity}
for all vectors $X,Y$. A metric satisfying the (skew)hermiticity conditions (\ref{hermiticity}) must have signature $(2n,2n)$ \cite{Dunajski:2006mk}.
Such a metric is referred to as \emph{neutral}.
On oriented 4-manifolds, every neutral hypercomplex structure allows a compatible hyperhermitian metric locally, 
(implicitly assumed in \cite{Dunajski:2001vt})
and globally after going to a double cover \cite{Grantcharov}.

Given a smooth oriented 4-manifold $\mathcal M$, there are two equivalent sufficient and necessary conditions for $\mathcal M$ to admit a neutral hypercomplex structure. 
\begin{enumerate}
	\item[a)] $\mathcal M$ admits two complex structures $J_+, J_-$ with the same orientation, such that $\{J_+, J_-\}=2c$ for $c$ constant with $|c|>1$ \cite{Lindstrom:2005zr}.
	\item[b)] $\mathcal M$ admits a basis of self-dual 2-forms $\Omega_I, \Omega_S, \Omega_T$ and a 1-form $\Theta$ (the \emph{Lee-form}) such that \cite{Dunajski:2001vt}, \cite{Boyer'88}
	\begin{equation}\label{omegas}
		d\Omega_i = \Theta \wedge \Omega_i, \quad i=I,S,T.
	\end{equation}
\end{enumerate}
The 2-forms are the fundamental forms associated to $(I,S,T)$. If $\Theta=0$, $\Omega_i$ are closed and define three symplectic forms, and the structure is called {\em neutral hyperk\"ahler} \cite{Kamada'99} or \emph{hypersymplectic} \cite{Hitchin'90}. Then there is a metric of signature $(2,2)$ 
and the Levi-Civita connection agrees with the Obata connection.
\section{Neutral hyperk\"ahler  and bi-hermitian geometry.}
The target space geometry described by the generalized K\"ahler potential $K(\bbX{L,R}, \bbXB{L,R})$ is generalized K\"ahler geometry for a positive definite metric $g$ \cite{Lindstrom:2005zr}\cite{Lindstrom:2007qf}\cite{Lindstrom:2007xv} with generalized hyperk\"ahler as a subclass.
When the metric is indefinite of signature $(n,n)$ the corresponding structures has been called generalized pseudo-K\"ahler and generalized pseudo-hyperk\"ahler \cite{Grantcharov}. 
In their bi-hermitian guise \cite{Gates:1984nk} these geometries may be given by the data $({\cal{M}}, g, J_\pm)$ supplemented by the integrability conditions
\ber
d^c\omega_++d^c\omega_-=0,~~dd^c\omega_\pm=0,
\eer{intbh}
where $\omega_\pm$ are the canonical two-forms associated with the two complex structures $J_\pm$. The integrability conditions imply the existence of a closed three-form $H$. Locally it may be written as $H=dB$ for some  two-form $B$. The three-form $H$ also enters the geometry as the torsion in the connections preserving $J_\pm$:
\ber
\nabla^\pm J_\pm=0, \quad \nabla^\pm =\nabla^0\pm \frac{1}{2}\,  g^{-1}H,
\eer{torsion}
where $\nabla^0$ is the Levi-Civita connection and (\ref{torsion}) is a consequence of (\ref{intbh}).

A condition which ensures (neutral) hyperk\"ahler geometry is 
\ber
\{J_+,J_-\}=2c\mathbb{I},
\eer{54}
where $c$ is a constant. The resulting geometry is radically different depending on whether $|c|>1$ or $|c|<1$. This may be understood from the following relation:
\ber
([J_+,J_-])^2=4(c^2-1),
\eer{56}
which makes 
\ber
J\equiv \frac 1 {\sqrt {1-c^2}}\left(J_- + c  J_+ \right)
\eer{57}
a complex structure when $c^2<1$ and
\ber
S\equiv \frac 1 {\sqrt {c^2-1}}\left(J_- +c  J_+ \right)
\eer{58}
a local product structure when  $c^2>1$.

In the first case
\ber
K\equiv\frac 1 {2\sqrt {1-c^2}} [J_+,J_-]=\frac 1 {2\sqrt {1-c^2}}g\Omega^{-1}
\eer{581}
is a third complex structure, and the set $(I\equiv J_+,J,K)$ generate $SU(2)$.
In the second case,
\ber
T\equiv \frac 1 {2\sqrt {c^2-1} }[J_+,J_-]=\frac 1 {2\sqrt {c^2-1}}g\Omega^{-1}
\eer{582}
is a second product structure and the set  $(I\equiv J_+,S,T)$ generate $SL(2,\mathbb{R})\cong Sp(2)$.

Below, we shall be interested in the second case, i.e. when $|c|> 1$, in which case the geometry corresponding to the set $(I,S,T)$ is called neutral hypercomplex, as reviewed in the previous section. 

When a bi-hermitian sigma model is written entirely in terms of semi-chiral fields, as in the case at hand (\ref{action}), the $B$-field is globally defined and (in a particular gauge) given by 
\ber
B=\Omega \left\{J_+,J_-\right\},
\eer{Bdef}
 where 
\ber
\Omega=\frac{1}{2}\left( \begin{array}{cc}
0&K_{LR}\cr
-K_{RL}&0\end{array}\right)
\eer{omega} 
is a symplectic form. Here $K_{LR}$ is defined in (\ref{Kmatrix}) and $K_{RL}$ is the transpose of  $K_{LR}$. Since we assume (\ref{54}), the right hand side in (\ref{Bdef}) is equal to $ 2\Omega c$ and hence $dB=H=0$. Further, the metric is  
\cite{Lindstrom:2005zr},\cite{Sevrin:1996jr}
\ber
g=\Omega [J_+,J_-]=2(\sqrt{c^2-1})\Omega T.
\eer{gdef}
The Lee-form $\Theta$ in 
(\ref{omegas}) vanishes and 
\ber
\Omega_T&=&2(\sqrt{c^2-1})\Omega,\cr
\Omega_I&=&2(\sqrt{c^2-1})\Omega S,\cr
\Omega_S&=&2(\sqrt{c^2-1})\Omega I.
\eer{forms}

The integrability conditions, which follow from (\ref{intbh}) are:
\beq
\nabla^0 I=0, \quad
\nabla^0 S =0, \quad
\nabla^0 T= 0,
\eeq{583}
where the connection is now torsion-free.
Note that this means that, for this case, the Obata connection (\ref{obata}), agrees with the Levi-Civita connection $\nabla^0$.
Also, the $c^2>1$ case implies that the metric is indefinite, in fact in the four-dimensional case at hand the signature is $(2,2)$, often referred to as the metric being neutral, as in section \ref{neutral_hypercomplex_structures} above.

In the four dimensional case we also have
\ber
J_+=\left( \begin{array}{cc}
\jmath&0\cr
K^{-1}_{RL} C_{LL}& K^{-1}_{RL}\jmath K_{LR}\end{array}\right),\quad
J_-=\left( \begin{array}{cc}
K^{-1}_{LR}\jmath K_{RL}& K^{-1}_{LR} C_{RR}\cr
0&\jmath\end{array}\right),
\eer{jpm}
where 
\ber
\jmath \equiv \left( \begin{array}{cc} 
i&0\cr
0&-i\end{array}\right), \quad C_{LL}= 2iK_{1\bar 1}\left( \begin{array}{cc} 
0&1\cr
-1&0\end{array}\right),
\quad C_{RR}= 2iK_{2\bar 2}\left( \begin{array}{cc} 
0&1\cr
-1&0\end{array}\right).
\eer{j}
Using this we find that (\ref{54}) is equivalent to
\ber
(1+c)|K_{12}|^2+(1-c)|K_{1\bar 2}|^2=2K_{1\bar 1}K_{2\bar 2}.
\eer{55}
This relation generalizes the Monge-Amp\` ere equation, and,  as described in the next section,  is satisfied by our solution (\ref{50}).
\section{A class of neutral hyperk\"ahler structures}
Combining the equations (\ref{invcndn}) and (\ref{55}) yields
\beq
	c=-\frac{\kappa \bar\kappa+1}{\kappa \bar\kappa-1}~.
\eeq{c}
Since $\kappa$ is non-zero, $|c| > 1$. The twisted supersymmetry transformations (\ref{lintfs}) break down at $|\kappa| \rightarrow \pm \infty$, which suppresses the limit $c\rightarrow -1$, i.e., the limit when the complex structures commute.

This means that we have found a two-parameter family of neutral hyperk\"ahler structures with a potential for the geometry given by (\ref{50}). The full set of geometric data, metric, $B$-field, complex structures and local product structures is expressible in terms of (functions of) the second derivatives of this potential. As for the positive definite case, it should be possible to linearize this structure and find the nonlinear structure as arising from a quotient, but we shall not pursue this issue here.

Using the relations (\ref{omega})-(\ref{j}) we find the following relations for the fundamental two-forms:
\ber
\Omega_T&=&(\sqrt{c^2-1})\left( \begin{array}{cc}
0&K_{LR}\cr
-K_{RL}&0\end{array}\right),\cr
&&~\cr
\Omega_I&=&\left( \begin{array}{cc}
cC_{LL} &c\jmath K_{LR}+K_{LR}\jmath\cr
-cK_{RL}\jmath-\jmath K_{RL}&-C_{RR}\end{array}\right),\cr
&&~\cr
\Omega_S&=&(\sqrt{c^2-1})\left( \begin{array}{cc}
C_{LL}&\jmath K_{LR}\cr
-K_{RL}\jmath&0\end{array}\right).
\eer{OmegaTSI}
 
As mentioned above, the generalized K\"ahler potential
has $c$ constant and  independent of the values of the parameters in $y$. This is seen 
by a direct calculation of the anticommutator in (\ref{Bdef}) \cite{Buscher:1987uw}. 
A constant $c$ implies from (\ref{Bdef}) that the torsion vanishes,
\beq
	dB = 2c \cdot d\Omega = 0.
\eeq{notorsion}

\subsection{Examples}
We now choose $\kappa =\sqrt{2}$, which implies $c=-3$ in (\ref{c}), and the parameters $\alpha, \beta, \gamma $ and $\delta$ in a way consistent with the conditions (\ref{51})\footnote{This represents a very particular choice with all the parameters real.}:
\beq
y=\bbX{L}-\nu\bbXB{L} +\bbX{R} +\nu\bbXB{R},
\eeq{specy}
with $\nu \equiv \sqrt {2}+1$.
For future use we note the split into real and imaginary part $y=\varphi+i\rho$ according to
\ber
2\varphi &=&\sqrt{2} \left( -(\bbX{L}+\bbXB{L})+\nu(\bbX{R}+\bbXB{R})\right),\cr
2i\rho&=&\sqrt{2} \left( \nu(\bbX{L}-\bbXB{L})-(\bbX{R}-\bbXB{R})\right).
\eer{realim}
The objects needed to find the fundamental two-forms are, according to (\ref{OmegaTSI})
\beq
K_{1\bar 1}=-\nu (F''+\bar F'') \equiv -\nu \Sigma=-K_{2\bar 2},
\eeq{exomega1}
\beq
K_{LR}=\nu\left(\begin{array}{cc}
\sqrt{2}D-\Sigma&D\cr
-D&-(\sqrt{2}D+\Sigma)\end{array}\right),
\eeq{exomega}
where $D\equiv F''-\bar F''$.
Note that the fundamental two-forms are all linear in the second derivatives of the potential $K$. This is true neither for the structures $I,S,T$ nor for the metric. Defining $E$ to be the sum of the metric and $B$-field, $E=g+B$,
and the submatrices according to
\beq
E=\left(
	\begin{array}{cc}
	E_{LL} & E_{LR} \\
	E_{RL} & E_{RR} \\
	\end{array}
	\right),
\eeq{Esub}
we may use the general formulae in \cite{Lindstrom:2005zr} to calculate
\ber
  E_{LL}&=&\frac{\nu\Sigma}{2|F''|^2}\left(
	\begin{array}{cc}
	2D(\Sigma-\sqrt{2}D) & 3D^2-\Sigma^2  \\
	3D^2-\Sigma^2& -2D(\Sigma+\sqrt{2}D)
	\end{array}\right),\cr
&&~\cr
E_{LR}&=&\frac{\nu}{4|F''|^2}\left(
	\begin{array}{cc}
	-(\sqrt{2}D-\Sigma)(5\Sigma^2-D^2) & -D(3\Sigma^2+D^2)  \\
	D(3\Sigma^2+D^2)&(\sqrt{2}D+\Sigma)(5\Sigma^2-D^2) 
	\end{array}\right),\cr
	&&~~\cr
	E_{RL}&=&\frac{\nu}{4|F''|^2}\left(
	\begin{array}{cc}
	(\sqrt{2}D-\Sigma)(\Sigma^2-5D^2) & -D(3\Sigma^2-7D^2)  \\
	D(3\Sigma^2-7D^2)&-(\sqrt{2}D+\Sigma)(\Sigma^2-5D^2) 
	\end{array}\right),\cr
	&&~\cr
	E_{RR} &=& \frac{\nu\Sigma}{2|F''|^2}\left(
	\begin{array}{cc}
	2D(\Sigma-\sqrt{2}D) & \Sigma^2-3D^2  \\
	\Sigma^2-3D^2 & -2D(\Sigma + \sqrt{2}D)
	\end{array}\right).
\eer{ELL}

It serves as a gratifying check on our algebra that the $B$-field calculated from (\ref{ELL}) is indeed
\beq
B=-6\Omega.
\eeq{Bcheck}
(Cf. (\ref{Bdef})).

To proceed, we need to choose a particular function $F(y)$. The simplest non-trivial  choice is the quadratic solution $F=y^2$ and corresponds to 
\beq
K=\textrm{Re} (\alpha\beta) \bbX{L}\bbXB{L}+ \textrm{Re}(\gamma\beta) \bbX{R}\bbXB{R} +\bigr[\alpha\gamma + \bar\beta\bar\delta \bigr] \bbX{L}\bbX{R} +
\bigr[\alpha\delta + \bar\beta \bar\gamma\bigr]\bbX{L}\bbXB{R}
+ c.c.,
\eeq{53b}
modulo terms that are killed by the integration measure ("K\"ahler gauge transformations").
It has $D=0$ and $\Sigma=4$ and describes flat space with metric
\beq
	g=8\nu \left(
	\begin{array}{cccc}
	0  & -1 & 1 & 0 \\
	-1 & 0 & 0 & 1 \\
	1 & 0 & 0 & 1\\
	0 & 1 & 1 & 0
	\end{array}\right).
\eeq{quadraticmetric}
The metric has the expected signature $(--++)$.

A more interesting example results from $F=e^{y}$. It has $D=2ie^{\varphi}\sin \rho$ and $\Sigma=2e^{\varphi}\cos \rho$. The metric will have an exponential pre-factor $e^\varphi$, the rest of the metric will depend on the angular variable $\rho$ only. 

We may impose further conditions. Requiring the metric to have real entries, e.g., implies $y=\varphi + i\pi n, n\in \mathbb Z$, and we recover the same metric as for the quadratic case (\ref{quadraticmetric}) but with $F''=e^\varphi$.

\section{Comments}
In this note we have investigated additional linear symmetries for a $N=(2,2)$ sigma model with semi-chiral fields restricted to four-dimensional target space. We have found that no solution for additional supersymmetry exists, but that a class of interesting solutions for $N=(4,4)$ twisted supersymmetry can be found. 

We have described how neutral hyperk\"ahler structures arise from a certain subclass of bi-hermitian geometries. In particular, they are  described locally by a single generalized potential. A similar statement can be found in \cite{Dunajski 2}, where Prop. 3  states that every hyperhermitian metric locally arises from a pair of complex valued functions satisfying a certain non-linear differential equation involving both first and second derivatives of the potentials. We have not  investigated the exact relation to our (real) generalized potential.

On compact complex surfaces, the neutral hyperk\"ahler structures have been classified \cite{Kamada'99}. They are 4-tori or Kodaira surfaces. An analysis of which functions $F(y)$ lead to compact complex surfaces is therefore of great interest.

It is further well-known that there exists an infinite family of $4d$ neutral hyperk\"ahler structures \cite{Grantcharov}.  In our examples  the arbitrariness in choice of function $F$ in (\ref{50}) again leads to an infinite family of such structures.
\vspace{1.5cm}

\noindent{\bf\Large Acknowledgement}:

\noindent
We thank Martin Ro\v cek for pointing out an omission in an earlier version of this paper and Gueo Grantcharov, Chris Hull, Rikard von Unge and Maxim Zabzine for discussions. 
We are grateful to the workshop ``Supersymmetry in complex geometry'' at the IPMU of the University of Tokyo, January 2009, for providing stimulating interactions between mathematicians and physicists.
The research of UL was supported by EU grant (Superstring theory)
MRTN-2004-512194 and VR grant 621-2006-3365.

\end{document}